\documentclass[usegraphicx,usenatbib,usedcolumn,useAMS]{mn2e}

\title[The sizes of intermediate-$z$ cluster discs]%
      {\vspace*{-0.5cm}The sizes of disc galaxies in intermediate-redshift clusters\vspace*{-0.5cm}}
\author[S. P. Bamford et al.]{%
  S. P. Bamford$^{1}$\thanks{E-mail: steven.bamford@port.ac.uk},
  B. Milvang-Jensen$^{2}$, A. Arag\'on-Salamanca$^{3}$\\
  $^{1}$ICG, University of Portsmouth,
  Mercantile House, Hampshire Terrace, Portsmouth, PO1 2EG, UK\\
  $^{2}$Dark Cosmology Centre, Juliane Maries Vej 30,
  2100 Copenhagen \O, Denmark\\
  $^{3}$School of Physics and Astronomy, University of Nottingham, 
  University Park, Nottingham, NG7 2RD, UK\\
}

\begin{document}
  
\date{Accepted ???. Received ???; in original form ???}

\pagerange{\pageref{firstpage}--\pageref{lastpage}} \pubyear{2007}

\maketitle

\label{firstpage}

\begin{abstract}
  We examine how the location of star formation within disc galaxies
  depends on environment at intermediate redshift.  This is achieved
  by comparing emission-line ($r_{\rmn{em}}$) and restframe $B$-band
  ($r_B$) scalelengths for matched samples of 50 field and 19
  cluster star-forming, disc galaxies, with $0.25 \leq z \le 1.0$ and
  $M_B \leq -19.5$ mag.  We find that at a given $r_B$ the
  majority of our cluster galaxies have $r_{\rmn{em}}$ smaller than those
  in the field, by 25 percent on average.  These results are compared
  with studies of local galaxies, which find a very similar behaviour.
  From the relations of $r_{\rmn{em}}$ and $r_B$ versus $B$-band
  absolute magnitude ($M_B$) we infer that the difference between the
  intermediate-$z$ cluster and field samples is mostly attributable to
  variation in $r_{\rmn{em}}$ at a given $M_B$, while the $r_B$ versus
  $M_B$ relation is similar for the two samples.

\end{abstract}



 \section{Introduction}

 The sizes of galaxy discs provide important constraints on a number of
 aspects of galaxy formation.  In the modern paradigm of cold dark
 matter \citep{1991ApJ...379...52W} and the halo model
 \citep{2002PhR...372....1C}, disc sizes are expected to be simply
 related to the properties of the dark matter halo each inhabits
 \citep{1998MNRAS.295..319M}.  Observed differences from this simple
 expectation provide insight into the details of disc galaxy formation
 and evolution \citep[e.g.,][]{1998MNRAS.297L..71M}.

 The surface brightness profiles of discs, as measured on broadband
 surface photometry, are ubiquitously found to have exponential
 profiles
 \citep{1970ApJ...160..811F,1981AA....95..105V,1982AA...110...61V}.
 The scalelengths of these profiles are thus a sufficient descripion of
 disc sizes (ignoring the truncation at large radii seen in many disc
 galaxies). These profiles reflect the stellar mass distribution in the
 discs, modulo variations due to radial trends in the age and
 metallicity of the stellar population and extinction.  Such trends are
 found in studies of local galaxies \citep{1996AA...313..377D}, however
 the variations within individual discs are generally small.


 One can also examine the distribution of emission-line flux in
 galaxies.  This may be achieved using narrow-band surface photometry
 or spatially resolved spectroscopy.  Emission line luminosity is
 proportional to the rate of ongoing star formation
 \citep[e.g.,][]{1998ARAA..36..189K}.  This is particularly true for the
 H$\alpha$ and H$\beta$ lines. The other strong lines are mainly due
 to forbidden transitions of oxygen, and have a metallicity dependence.
 While this is fairly weak, it is possible that radial metallicity
 trends \citep[e.g.,][]{1995ApJ...438..170H} play a role in the
 profiles of the forbidden lines.  Herein we assume emission-line flux
 profiles represent the variation in star formation rate with radius.
 These profiles are also generally found to be exponential (at least
 outside of the region dominated by any bulge component), and thus can
 be well described by a scalelength, $r_{\rmn{em}}$.

 It has long been known that galaxies in clusters have distributions of
 properties different from those in the field, in terms of their
 stellar populations \citep{1985ApJ...288..481D}.  and morphologies
 \citep{1980ApJ...236..351D}.  While many of these differences may be
 an effect of the biased distribution of parent halo masses in dense
 regions, it appears that at least some aspects are due to rather more
 recent physical interactions between galaxies and their wider
 environment.  The evolution of cluster galaxy populations with cosmic
 time has been directly observed, from star formation properties
 \citep[e.g.,][]{1999ApJ...518..576P,2006ApJ...642..188P} and morphologies
 \citep[e.g.,][]{1997ApJ...490..577D}.  A wealth of studies thus
 suggest that a substantial fraction of galaxies have transformed from
 star-forming to passive over the last half of the age of the universe,
 with such transformations being more frequent in dense regions.  While
 some transformations may also transform disc galaxies to ellipticals,
 the existence of passive spirals \citep{1998ApJ...497L..75K}, and
 prevalence of lenticular galaxies \citep{1980ApJ...236..351D}, both
 preferentially found in clusters, are evidence that many do not
 involve major mergers.  Furthermore, these suggest that the transition
 is fast ($\sim 1$ Gyr); a sudden truncation of star-formation rather
 than a gradual decline.  However, the importance of the various
 potential transformation mechanisms remains unclear, as is the density
 at which such environmental effects begin to dominate.

 The picture of a rapid transformation from star-forming to passive is
 supported by studies of the distribution of colour and emission-line
 strength of nearby galaxies as a function of local galaxy density.  In
 both quantities two distinct populations, star-forming and passive,
 can easily be identified.  However, while the relative proportions
 vary strongly with environment, the distributions of the individual
 components do not change significantly, and there is little evidence
 for an emerging transition population
 \citep{2004MNRAS.348.1355B,2006MNRAS.373..469B}.  These studies
 suggest that the properties of \emph{star-forming} galaxies do not
 change with environment.

 There is, however, evidence for some differences between star-forming
 galaxies in clusters and in the field.  These include the findings of
 \citet{2006AJ....131..716K}, that the ratio of emission-line to
 broadband scale length differs significantly between galaxies in
 clusters and the field.  This is a reflection of the result of
 \citet{2004ApJ...613..851K} that, on average, there is a suppression
 of star-formation in the outer regions and some enhancement in the
 inner regions of cluster galaxies.  \citet{2000MNRAS.317..667M} also
 show that the fraction of disc galaxies with compact, central
 star-formation increases with both local density and the central
 density of any encompassing cluster.  \citet{2004ApJ...613..851K}
 find that cluster disc galaxies have median star formation rates
 lower than the field by a factor of two.  The opposing finding, a
 lack of significant variation in the star-forming galaxy H$\alpha$ EW
 distribution, by \citet{2004MNRAS.348.1355B} may be because their
 3-arcsec fibre spectra do not sample the outer regions of galaxies.

 It remains unclear if these limited differences are intrinsic, or
 whether recent environmental effects are responsible.  However, strong
 support for an environmental interaction in local clusters,
 particularly with the intra-cluster medium, is supplied by the
 enhanced fraction of galaxies with truncated H$\alpha$ emission
 \citep{2004ApJ...613..866K}, deficient or truncated HI discs
 \citep{2004AJ....127.3300V}, and direct observations of ongoing gas
 stripping \citep{2005AJ....130...65C}.

 While recently work has begun to constrain the variation in stellar
 scalelengths at intermediate redshift
 \citep[e.g.,][]{2005ApJ...635..959B,2006ApJ...650...18T}, the
 variation of star-formation scalelength has been left largely
 unstudied.  An indication of a difference in the ratio of
 star-formation to stellar scalelength between galaxies in one $z \sim
 0.83$ cluster and those in the coeval field was found by
 \citet{2003PhDT.........1M}, using an earlier analysis of part of the
 data considered in this letter.  Beyond this, there has been no
 attempt to examine evolution in the dependence of the ratio on
 environment.  We consider this further here, using a sample five times
 larger than previously.

 \section{Data}

 \begin{figure}
 \centering
 \includegraphics[height=0.32\textwidth,angle=270]%
                 {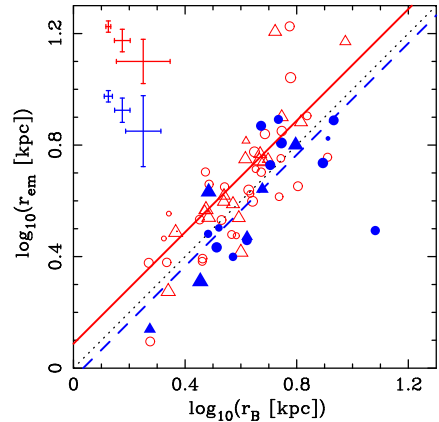}
                 \caption{\label{fig:rd} A plot of $\log_{10}(r_{\rmn{em}})$
                   versus $\log_{10}(r_B)$ for galaxies in the
                   matched samples. Field galaxies are shown by open,
                   red symbols, cluster galaxies by filled, blue
                   symbols.  Triangles and circles indicate $r_B$
                   was measured on HST- or ground-based imaging,
                   respectively. The error bars in the top left corner
                   give the 10th-, 50th- and 90th-percentiles of the
                   distribution of uncertainties on the plotted field
                   and cluster points.  The size of each point is
                   linearly related to the error on that point, with
                   larger points having smaller uncertainties.  The
                   dotted line traces the one-to-one relation, while
                   the solid and dashed lines indicate fixed-slope fits
                   to the field and cluster points.}
 \end{figure}

 \begin{figure}
 \centering
 \includegraphics[height=0.32\textwidth,angle=270]%
                 {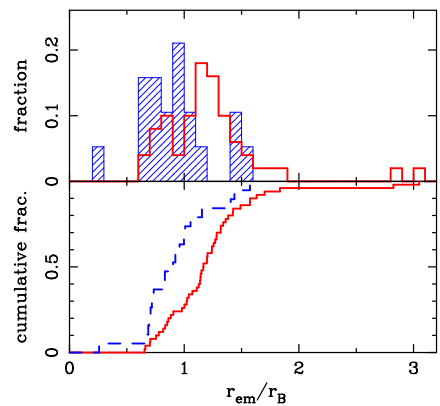}
                 \caption{\label{fig:rd_ratio_distrib_ks} Distribution
                   (top) and cumulative histogram (bottom) of
                   $r_{\rmn{em}}/r_B$.  The field sample is shown by the
                   red, solid line, while the hatched blue area
                   and dashed line indicate the cluster sample.}
 \end{figure}

 We employ VLT/FORS2 \citep{1998Msngr..94....1A} multi-slit
 spectroscopy of disc galaxies in the fields of five massive clusters
 at $0.2 \la z \la 0.8$, supplemented by broadband ground-based and
 Hubble Space Telescope (HST) imaging.  The clusters studied are
 MS0440, AC114, A370, CL0054 and MS1054, with velocity dispersions
 $\sigma_{\rmn{cl}} \sim 740$--$1400$ kms$^{-1}$
 \citep{2001ApJ...548...79G}.  The target selection is based on a
 number of aspects related to the likelihood of obtaining emission-line
 observations suitable for the reliable measurement of rotation
 velocities and scalelengths. Given the differing prior information for
 each field the selection is thus somewhat heterogeneous.  However,
 while known cluster galaxies were favoured, there were no other
 differences in the selection criteria of cluster and field galaxies.
 The observations therefore provide a basis for fair comparison between
 the two samples.

 From these data we have constructed samples of 19 cluster and 50 field
 galaxies, with matched redshift ($0.25 \leq z \le 1.0$) and luminosity
 ($M_B \leq -19.5$ mag) ranges, and with reliable emission-line
 ($r_{\rmn{em}}$) and broadband ($r_B$) scalelengths.  The data and
 construction of the matched samples are described fully in
 \citet{2005MNRAS.361..109B}, where they are used to investigate
 offsets of the Tully-Fisher relation between clusters and the field,
 following a pilot study on an earlier analysis of the MS1054 data
 alone \citep{2003MNRAS.339L...1M}. Complete data tables, and an earlier
 version of the analysis in this letter, can be found in
 \citet{2006PhDT.........5B}.  These data have also been used to
 investigate differences in the chemical and star-formation properties
 of intermediate-$z$ cluster and field galaxies
 \citep{2006MNRAS.368.1871M}, as well as the redshift evolution of the
 Tully-Fisher relation \citep{2006MNRAS.366..308B} and
 luminosity--metallicity relation \citep{2006MNRAS.369..891M}.

 The emission-lines in the 2-dimensional spectra were analysed by a
 method dubbed \textsc{elfit2py}, whereby model emission lines are
 iteratively fit to the data to determine the parameter set which best
 describes the data, in a manner based on the technique of
 \citet{1999PASP..111..453S}.  Several quality criteria, including
 visual inspection, were used to reject any fits deemed unreliable.
 The primary quantities thus obtained are the rotation velocity and
 scalelength of the emission, which is assumed to follow an
 exponential profile.  Note that this method differs from that
 employed by most other groups \citep[e.g.,][]{2004AA...420...97B} in
 that we directly compare 2-d spectra of the real and model emission
 lines, rather than their traces.  In this way we can constrain the
 emission scalelength.  As the primary purpose is usually to determine
 the rotation velocity, most other methods hold the emission-line
 scalelength fixed at some factor of the broadband scalelength.  Note
 that the $r_{\rmn{em}}$ we obtain by this method are corrected for
 the effect of seeing.  The emission lines measured are [OII]$\lambda
 3727$, H$\beta$, [OIII]$\lambda4959$ and [OIII]$\lambda5007$. Usually
 more than one emission line was reliably fit for each galaxy, in
 which case we take the weighted mean of their parameters, after
 checking for consistency.

 Absolute restframe magnitudes $B$-band were obtained from
 \textsc{sextractor} \citep{1996AAS..117..393B} Kron magnitudes, and
 k-corrected using available colour information to constrain the SED
 employed for each galaxy.  Note that the magnitudes used in this
 letter do not include a correction for internal extinction (contrary
 to previous papers using these data), to aid comparison with other
 studies.

 Seeing-corrected, broadband, disc scalelengths were obtained by
 fitting bulge + disc models to the ground-based and HST imaging using
 \textsc{GIM2D} \citep{2002ApJS..142....1S}.  Only bands close to
 $R$-band were used, ground-based $R$ and HST $F606W$, $F675W$ or
 $F702W$, corresponding roughly to restframe $B$-band at the median
 redshift of our samples, $z \sim 0.5$.  HST measurements were
 preferred, but for 62 percent of our sample only ground-based
 measurements were available.  We have checked that the conclusions of
 this letter are unaffected by using HST- or ground-based measurements
 alone, except for a lower confidence due to reduced sample sizes.

 Throughout this letter absolute magnitudes and scalelengths are given
 for a Friedmann-Robertson-Walker cosmology with $\Omega_{\Lambda} =
 0.7$, $\Omega_m = 0.3$ and $H_0 = 70$~kms$^{-1}$~Mpc$^{-1}$.

 \section{Star-formation versus stellar scalelengths}

 \begin{figure}
 \centering
 \includegraphics[height=0.32\textwidth,angle=270]%
                 {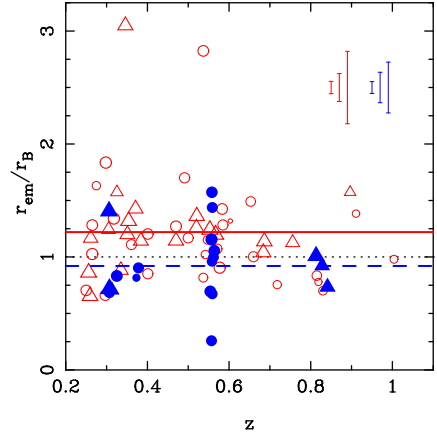}
                 \caption{\label{fig:rd_ratio_z} The emission-line
                   versus broadband scalelengths ratio,
                   $r_{\rmn{em}}/r_B$, versus redshift. Lines and
                   symbols are as in Fig.~\ref{fig:rd}.}
 \end{figure}

 In Fig.~\ref{fig:rd} we plot (using log axes) the scalelengths of
 emission-line flux versus broadband light for our matched samples of
 intermediate-$z$ cluster and field disc galaxies.  The slopes of
 the relation for each sample are not particularly well constrained,
 but show no evidence ($<1.4\sigma$) of differing from unity.  This is
 consistent with the evidence and usual assumptions at low-$z$.
 We therefore assume unit slope, i.e. that $r_{\rmn{em}}/r_B$
 does not vary with either quantity.  We can thus simply compare the
 distributions of this ratio.

 The distributions of $r_{\rmn{em}}/r_B$, normalised to unit
 area, are plotted in the top panel of
 Fig.~\ref{fig:rd_ratio_distrib_ks}. The distributions appear
 reasonably Normal given the limited sample sizes.  We estimate the
 average $r_{\rmn{em}}/r_B$ for each sample by simply taking the mean,
 and compute the uncertainty on the mean by the usual standard error.
 We also assessed the means and their uncertainites by bootstrap
 resampling, which gives practically identical results, further
 supporting the Normality of the distributions.  We find
 $r_{\rmn{em}}/r_B\,(\rmn{field}) = 1.22 \pm 0.06$ and
 $r_{\rmn{em}}/r_B\,(\rmn{cluster}) = 0.92 \pm 0.07$.

 To determine the significance of the difference between the sample
 means we employ Student's $t$-test.  Any process which may change
 $r_{\rmn{em}}/r_B$ may also affect the intrinsic scatter on
 $r_{\rmn{em}}/r_B$, even though the measurement uncertainties on the
 two samples are similar.  We assess this using an $F$-test, which
 gives the probability of both samples being drawn from Normal
 populations with the same variance as $8$ percent.  While not highly
 significant, unlikely given the small sample size, it suggests there
 may be a difference, in the sense that the cluster sample distribution
 shows a lower scatter.  We thus apply a $t$-test without the
 assumption of equal variances, finding a probability that the two
 samples have the same mean of $0.3$ percent, corresponding to a
 $3\sigma$ significance.  Excluding potential outliers, by imposing
 cuts of $r_B, r_{\rmn{em}} < 10$ kpc, does not change our overall
 conclusions.

 We can also examine the difference between the field and cluster
 samples through their cumulative histograms, as shown in the bottom
 panel of Fig.~\ref{fig:rd_ratio_distrib_ks}.  A K-S test gives the
 probability of the two samples being drawn from the same parent
 distribution as $0.4$ percent.

 The data therefore suggest that the galaxies in our cluster sample
 have, on average, significantly smaller $r_{\rmn{em}}/r_B$ values than
 the field galaxies.  These two samples of disc galaxies have been
 selected, observed and analysed together in the same way and have been
 matched in terms of redshift and luminosity. We thus conclude that
 bright, disc galaxies in intermediate-$z$ clusters possess
 significantly smaller ratios of their emission-line to broadband
 scalelengths than comparable field galaxies.

 As an aside, note that the finding that cluster and field galaxies
 have different ratios of the emission-line versus broadband
 scalelengths, both locally and at intermediate-$z$, supports the
 decision of \citet{2005MNRAS.361..109B} to allow this quantity to
 vary in their emission line fits.  As the adopted scalelength affects
 the weighting of emission from regions with different line-of-sight
 velocities, the use of a fixed ratio by other groups may cause
 systematic differences in the measured rotation velocities of cluster
 and field galaxies.

 Our findings at intermediate-$z$ correspond to a very similar
 result observed locally. \citet{2006AJ....131..716K} measure the ratio
 of H$\alpha$ emission-line ($r_{\rmn{H}\alpha}$) to $R$-band ($r_R$)
 scalelengths for galaxies in the nearby Virgo cluster
 ($\sigma_{\rmn{cl}} \sim 650$ kms$^{-1}$,
 \citealt{1996ApJ...457...61G}) and local field.  As our
 intermediate-$z$ measurements will be rather radially limited, we
 can best compare with their results fit over $1$--$2r_{R}$.  They find
 $r_{\rmn{H}\alpha}/r_R = 1.18 \pm 0.10$ and $0.91 \pm 0.05$ for field
 and cluster galaxies, respectively.  Remarkably these are very similar
 to our intermediate-$z$ findings, particularly within the
 uncertainties.  Note that we expect their $r_{\rmn{H}\alpha}$ to be
 equivalent to our $r_{\rmn{em}}$.  Their broadband scalelengths are
 measured in the $R$-band, while ours are approximately restframe
 $B$-band.  However, \citet{2005ApJ...635..959B} find that for local
 field galaxies the average change in disc scalelengths in going from
 $B$ to $R$-band is only a $\sim 3$ percent decrease.

 Finally, as our samples span a range in redshift, we can ask whether
 we can detect any significant evolution within our samples.  The
 ratio $r_{\rmn{em}}/r_B$ is plotted versus redshift in
 Fig.~\ref{fig:rd_ratio_z}.  No evolution is discernable given the
 scatter and small sample sizes.  With the uncertainties on our
 intermediate-$z$ measurements and those at low redshift, as well as
 the use of different techniques, we cannot quantify any evolution in
 the emission-line to broadband scalelength ratios for cluster and
 field galaxies between $z \sim 0.5$ and today.  However, we have
 presented strong evidence that very similar general behaviour is seen
 at both epochs.

 \section{Scalelengths versus luminosity}

 We now turn our attention to determining the cause of the difference
 in $r_{\rmn{em}}/r_B$ for the cluster and field disc galaxies in our
 intermediate-$z$ sample.  To do this we consider the relations of
 $r_B$ and $r_{\rmn{em}}$ versus $M_B$.  This also has the advantage of
 removing any effect due to differences in the luminosity functions of
 the cluster and field samples on the result of the previous section.

 First we require a well determined, comparable, local relation, at
 least for $r_B$ versus $M_B$.  To construct this we use data from the
 Millenium Galaxy Catalogue
 \citep[MGC:][]{2003MNRAS.344..307L,2005MNRAS.360...81D}.  This is a
 modern, large survey with high redshift completeness and good quality
 $B$-band imaging, complemented by a wealth of additional data from
 SDSS and 2dFGRS.  \citet{2006MNRAS.371....2A} performed bulge+disc
 fits to $\sim 10000$ galaxies in this survey, using GIM2D on $B$-band
 images.  These data are ideal for comparison with our GIM2D fits to
 restframe $B$-band imaging of intermediate-$z$ galaxies.  As with our
 data, the MGC magnitudes are uncorrected for internal extinction.
 The quantities have been converted to our adopted cosmology.  Note
 that evolution of the luminosity--size relation has been investigated
 using the MGC to define a local relation by
 \citet{2007astro.ph..1419C}, but here we specifically consider disc
 scalelengths.

\begin{figure}
 \centering
 \includegraphics[height=0.29\textwidth,angle=270]%
                 {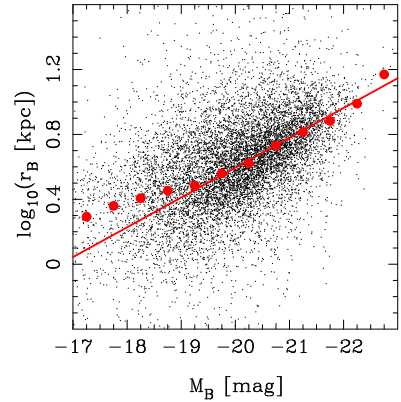}
                 \caption{\label{fig:ms_mgc} The $r_B$ versus $M_B$
                   relation for the Millennium Galaxy Catalogue. The
                   small points show individual galaxies, while the
                   large filled circles indicate the median scalelength
                   in bins of 0.5 in $M_B$. The solid line
                   is a fit to these median points.}
 \end{figure}

 Figure~\ref{fig:ms_mgc} plots the logarithm of disc $B$-band
 scalelength versus absolute $B$-band magnitude for the disc components
 of MGC galaxies.  Only objects judged by their quality control
 procedure to be physically real discs are included.  In order to fit a
 relation to these data, we have first plotted the median
 $\log_{10}(r_B)$ in bins of width 0.5 in $M_B$, and then fit a
 straight line to these points over the magnitude range we are
 concerned with, $M_B < -19$.  The resulting fit is $\log_{10}(r_B)
 = -0.184 M_B - 3.081$.  This line is very close to the relation
 obtained using the data from \citet{1995PhDT.......129D}, in which the
 scalelengths of nearby galaxy discs were measured using a similar
 technique.

\begin{figure}
 \centering
 \includegraphics[height=0.29\textwidth,angle=270]%
                 {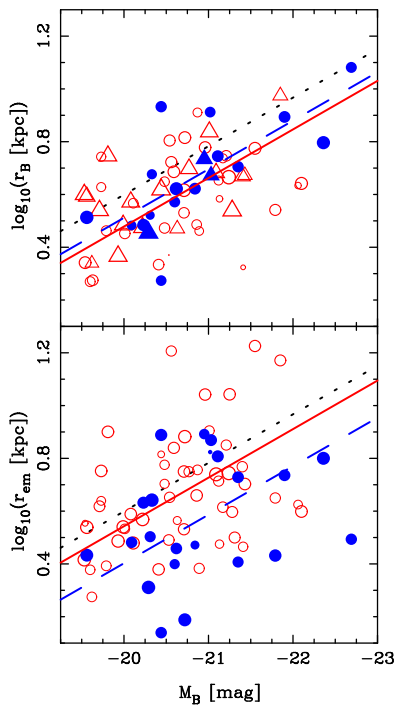}
                 \caption{\label{fig:ms_mrd} Relations between the
                   scalelengths, $r_B$ and $r_{\rmn{em}}$, and $M_B$.
                   Lines and symbols are as in Fig.~\ref{fig:rd}. The
                   dotted black line shows the local MGC
                   $r_B$--$M_B$ relation (identical in both panels).}
 \end{figure}

 The top panel of Fig.~\ref{fig:ms_mrd} shows the relation between
 $\log_{10}(r_B)$ and $M_B$ for our samples of intermediate-$z$
 field and cluster galaxies, along with the local MGC
 relation.  As the slope of free fits to the intermediate-$z$ data
 are ill-constrained we fix the slope to that of the local MGC data and
 fit each sample by taking the mean of the residuals from the MGC
 relation.  We obtain offsets from the MGC relation of $-0.12 \pm 0.02$
 and $-0.09 \pm 0.03$ for the cluster and field samples, respectively.
 The $z \sim 0.5$ field relation is thus different to that locally at a
 $\ga 5 \sigma$ confidence level, with the distant galaxies $\sim 25$
 percent smaller at a given $M_B$.  Though we have not given detailed
 consideration to selection effects, this agrees well with the results
 of dedicated studies, such as \citet{2005ApJ...635..959B}.  There is
 no significant difference between our intermediate-$z$ cluster
 and field samples ($< 1\sigma$).

 The lower panel of Fig.~\ref{fig:ms_mrd} shows 
 $\log_{10}(r_{\rmn{em}})$ versus $M_B$ for our intermediate-$z$ samples.
 The local MGC $\log_{10}(r_B)$ versus $M_B$ relation is reproduced
 for reference, and we assume the same slope for fitting the
 $\log_{10}(r_{\rmn{em}})$ versus $M_B$ relations.  As expected from the
 above findings, the cluster and field samples are offset from one
 another in this plot.  The separation in $\log_{10}(r_{\rmn{em}})$ is
 0.14 dex, with cluster galaxies having $28$ percent smaller
 $r_{\rmn{em}}$ than field galaxies of the same luminosity.  An
 $F$-test shows no evidence for differing variances, so we apply
 a $t$-test assuming equal variances, determining the significance of
 the offset between our intermediate-$z$ cluster and field samples as
 $98.9$ percent, corresponding to $2.5\sigma$ confidence.  Thus nearly
 all of the difference in $r_{\rmn{em}}/r_B$ between the
 intermediate-$z$ cluster and field samples can be attributed to
 differences in $r_{\rmn{em}}$ at a given $M_B$.

 \section{Conclusions}

 We find that, for a given broadband scalelength, cluster galaxies have
 emission-line scalelengths on average 25 percent smaller than those in
 the field at intermediate redshifts.  This implies that star-formation
 in the majority of these star-forming, disc, cluster galaxies is
 significantly more centrally concentrated, with respect to the stellar
 disc, than in similar field galaxies.  Our findings extend to
 intermediate redshift the behaviour seen in local studies.  We do not
 detect any significant redshift evolution in $r_{\rmn{em}}/r_B$ for
 cluster or field galaxies.

 From the relations of $r_{\rmn{em}}$ and $r_B$ versus $B$-band absolute
 magnitude, we infer that the difference between the intermediate-$z$
 cluster and field samples is mostly attributable to variation in
 $r_{\rmn{em}}$ at a given $M_B$.

 \section*{Acknowledgments}
 This work was based on observations made with ESO Telescopes at
 Paranal Observatory under programme IDs 066.A-0376 and 069.A-0312, and
 on observations made with the NASA/ESA Hubble Space Telescope,
 obtained from the data archive at the Space Telescope Institute.
 We have made use of the Millenium Galaxy Catalogue, 
 publicly available from \verb!http://www.eso.org/~jliske/mgc!.


 \bsp


{\small%

}

\label{lastpage}

\end{document}